\documentclass[aps,pre,nofootinbib,showpacs,twocolumn]{revtex4}%
\usepackage{amssymb,latexsym,mathrsfs}
\usepackage{amsfonts}
\usepackage{graphicx}

\newcommand{\beq}{\begin{equation}}
\newcommand{\eeq}{\end{equation}}
\newcommand{\beqn}{\begin{eqnarray}}
\newcommand{\eeqn}{\end{eqnarray}}
\newcommand{\bearr}{\begin{array}}
\newcommand{\enarr}{\end{array}}

\newcommand{\ket}[1]{|#1\rangle}
\newcommand{\bra}[1]{\langle#1|}

\newcommand{\ta}{\sigma}
\def\bea{\begin{eqnarray}}
\def\eea{\end{eqnarray}}
\def\ba{\begin{array}}
\def\ea{\end{array}}
\def\n{\nonumber}
\def\c{\mathscr}
\def\ket{\rangle}
\def\bra{\langle}
\def\c{\mathscr}
\begin{document}
\title{ Active Absorbing State Phase Transition Beyond Directed Percolation : A Class of Exactly Solvable Models}
\author{ Urna Basu}
\email[E-mail address: ]{urna.basu@saha.ac.in}
\author{P. K. Mohanty}
\email[E-mail address: ]{pk.mohanty@saha.ac.in}
\affiliation{Theoretical Condensed Matter Physics Division, Saha Institute of Nuclear Physics,
1/AF Bidhan Nagar, Kolkata, 700064 India.}
\date{\today}
\vskip 2.cm
\begin{abstract}
 
We introduce and solve a model of hardcore particles on a one dimensional 
periodic lattice which undergoes an active-absorbing state phase transition
at finite density. In this model an occupied site is defined to be \textit{active} 
if its left neighbour is occupied and the right neighbour is vacant. Particles 
from such active sites hop stochastically to their right. We show that, both the 
density of active sites and the survival probability vanish as the particle
density is decreased below half. The critical exponents and spatial 
correlations  of the model are calculated exactly using the matrix product 
ansatz. Exact analytical study of several variations of the model reveals 
that these non-equilibrium phase transitions  belong  to a new universality 
class  different from the generic active-absorbing-state phase transition,  
namely directed percolation.
\end{abstract}
\maketitle

One  of the most studied models of non-equilibrium phase transition is  
directed percolation(DP)\cite{dp} defined on a $d$-dimensional lattice, 
where an infected site stochastically infects its neighbours in one 
particular direction. Depending on the infection probability $p$,  
the infection may eventually 
survive (when $p>p_c$) or decay into the absorbing state  where no site is 
infected. Non-equilibrium phase transitions in 
several other systems, such as reaction diffusion systems\cite{react-diff}, 
depining transitions\cite{depinning}, 
damage spreading\cite{spread}, synchronization transition\cite{synchro}, 
sand-pile models\cite{pk},  and certain probabilistic cellular automata\cite{ca}, 
are  known to be in the universality class of DP.  It has been conjectured
\cite{gras} that an ``active-absorbing phase transition  governed 
by a  fluctuating scalar order parameter" generically belongs to the universality 
class of DP.  
There are certain exceptions, though. 
Particle-hole symmetry\cite{cdp}, conservation of parity\cite{parity}, and symmetry between 
different absorbing states\cite{sym} lead to different universalities. 
Again in sandpile models\cite{sp}, coupling  of  the order parameter to  
the conserving height fields\cite{cons-h} results in different critical 
behaviour. Also, conserved lattice gas (CLG) models\cite{clg, Oliveira} where the activity field is 
coupled to the conserved density show critical behaviour different from DP. 
This absorbing state phase transition in the presence of conserved field 
is not well understood and most studies in this direction are numerical.

In this paper, we provide an exact analytical solution for 
a  model  of hardcore particles  on a one dimensional 
ring which undergoes an  active-absorbing phase transition as
the density of particles is changed. The model is defined with a 
dynamics where  a particle from an occupied 
site hops to the right neighbouring site if the  left one is occupied.  
This restricted asymmetric exclusion process (RASEP)
leads to  a transition  from an active phase to an absorbing state 
as the density  of the system falls below $\frac 1 2$. 
The critical exponents of the system at the 
transition point and spatial correlations have been calculated  exactly using the matrix 
product ansatz(MPA)\cite{Derrida,mpa}. 
Some variations of the model, where  particles may hop to both 
directions stochastically,  or  hop to the right (left) only when it is 
followed by $\mu$ or more particles from left(right), could also be solved exactly.
These models, which  have same exponents at the  transition point,  form a new 
universality class of  active-absorbing phase transition different from the 
generic universality  class, namely, DP.

  The model is defined  on a one dimensional lattice labeled by 
sites $i=1,2\dots L$ which are either vacant or  occupied with at most one 
particle; corresponding site variables are taken $s_i=0,1$.  A periodic boundary 
condition is imposed so that $s_{i+L}= s_i$.  The dynamics of the system can be 
described as follows.  
A particle from a randomly chosen  site $i$ is  transferred to the right only 
if  $s_{i+1}=0$ and  $s_{i-1}=1$. This particle conserving dynamics is 
thus equivalent to a reaction diffusion system 
\begin{equation}
110 \to 101.
\label{eq:110}
\end{equation}
We define the activity 
field at site $i$ as $\phi_i=  s_{i-1}s_i(1-s_{i+1})$ which  
takes values $1$ or $0$ depending on whether the 
site $i$ is active, $i.e.,~ s_i=1=s_{i-1} ~ {\rm  and}  s_{i+1}=0$. 
The density of active sites
\beq
\bra \phi_i\ket= \langle s_{i-1} s_{i} (1-s_{i+1}) \rangle
\label{eq:rhoa}
\eeq
is denoted by $\rho_a$ in the thermodynamic limit. 
A configuration is said to be active  if  there is  at least one active site, 
otherwise it is called absorbing. 
For a system of $N=\sum_i^L s_i$ particles density is $\rho= \frac N L$. Clearly, 
there is  only one configuration at $\rho=0$ (and at $\rho=1$)  which is absorbing. 
First let us consider the regime $\rho\le \frac 12$ where there are both active and absorbing configurations. 
Total number of  absorbing  configurations in this regime is  $\frac{L}{L-N} C^{L-N}_N$
and  the rest are active. In this regime the system is arrested in one of these absorbing 
configurations  in steady state resulting in $\rho_a=0$.
For $\rho> \frac12$, however, there is  no absorbing configuration
which corresponds to an active phase  with 
fluctuating density of active sites. Thus, $\rho_a$ can be  taken as the 
order parameter of this active absorbing state  phase transition occurring at 
$\rho_c=\frac 12$. 
\begin{figure}
\includegraphics[width=7cm]{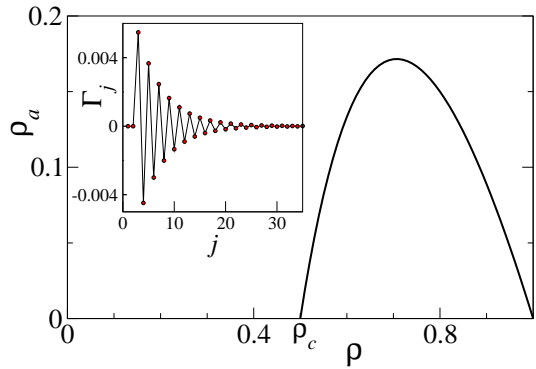}
\caption{ The order parameter $\rho_a$ for  RASEP is 
nonzero for $\rho>1/2$. Inset shows the decay of correlation function 
$\Gamma_j$ defined in Eq.(\ref{eq:gamma}).}
\end{figure}

  One can describe the dynamics of the model  alternatively in terms of  the bond  
variables $\tau_i$, which connects the sites $s_i$ and $s_{i+1}$. Correspondingly, 
we choose $\tau_i = 2s_i+s_{i+1}$ for four possible combinations $(s_i,s_{i+1})$.
Note that every configuration $\{s_i\}$ can be uniquely translated to $\{\tau_i\}$ and 
vice verse. The dynamics can be described in terms of $\tau_i$ as a reaction system 
\beq
321\rightarrow 213, ~~~~~~320\rightarrow 212.
\label{eq:321}
\eeq
The fact that $\tau_i$ and $\tau_{i+1}$ have  a common  site $s_{i+1}$ 
puts certain restrictions  on the allowed  configurations. However one 
need not bother about those as any 
configuration translated from $\{s_i\}$ automatically 
satisfies these restrictions and the dynamics (\ref{eq:321}) respects the same.
 
    To  get a steady state distribution for the reaction system, either 
Eq.(\ref{eq:110}) or Eq.(\ref{eq:321}) which has  three site dynamics,  we 
generalize the formulation  of  matrix product ansatz \cite{Derrida,mpa} which is commonly used 
for a two site dynamics. This generalization is different from what has been discussed earlier\cite{Andreas}. 
Let us describe the formulation in generic terms before using 
it in this specific problem. In MPA, first a configuration $\{\sigma_1, \sigma_2,\dots 
\sigma_L\}$ is translated to a  product of matrices   
 by  replacing each 
$\sigma_i$ by a  matrix  $A_{\sigma_i}$ and an ansatz is made that for a periodic system,
the unnormalized weight in the steady state is  given by  
\beq
f(\sigma_1, \sigma_2,\dots \sigma_L) = Tr[A_{\sigma_1}, A_{\sigma_2},\dots A_{\sigma_L}].
\label{eq:mpa}
\eeq

      This  ansatz could provide an exact solution  for any three site dynamics 
if one can find matrices $A_\sigma$ such that steady state weight (\ref{eq:mpa}) 
satisfies the corresponding master equation in steady state,

\vspace*{0.1cm}
\bea
0&=&\frac{d}{dt} f(\ta_{1}\ta_{2}...\ta_{L})  \cr 
&= &\sum_{i,\ta^\prime}W(\ta_i^\prime\ta_{i+1}^\prime\ta_{i+2}^\prime \rightarrow \ta_i\ta_{i+1}\ta_{i+2})f(...\ta_{i}^\prime\ta_{i+1}^\prime\ta_{i+2}^\prime...) \cr
&-&\sum_{i,\ta^\prime}W(\ta_i\ta_{i+1}\ta_{i+2} \rightarrow \ta^\prime_i\ta^\prime_{i+1}\ta^\prime_{i+2})f(...\ta_{i}\ta_{i+1}\ta_{i+2}...). \nonumber
\eea
Here $W$s are the transition rates for  the three site dynamics. 
Right hand side of  the master equation can 
be  arranged to vanish  for any generic three site dynamics if
\bea
&&\sum_{\ta^\prime}W(\ta_i^\prime\ta_{i+1}^\prime\ta_{i+2}^\prime \rightarrow \ta_i\ta_{i+1}\tau_{i+2})f(...\ta_{i}^\prime\ta_{i+1}^\prime\ta_{i+2}^\prime...) \cr
&-&\sum_{\ta^\prime}W(\ta_i\ta_{i+1}\ta_{i+2} \rightarrow \ta^\prime_i\ta^\prime_{i+1}\ta^\prime_{i+2})f(...\ta_{i}\ta_{i+1}\ta_{i+2}...)\cr
&=&Tr[...\tilde A_{\ta_i}\tilde A_{\ta_{i+1}}A_{\ta{i+2}}...]- Tr[...A_{\ta_i}\tilde A_{\ta_{i+1}}\tilde A_{\ta_{i+2}}...]
\label{eq:cancel}
\eea
where $\tilde A_\sigma$ are  auxiliary matrices. Equation  (\ref{eq:cancel}) is a
generalization of the cancellation procedure introduced earlier \cite{Derrida} 
for two site dynamics. Such a cancellation  is feasible only when 
one can find matrices and auxiliaries  which satisfy  Eq.(\ref{eq:cancel}) for a 
specific dynamics. 

Now let us try to apply this generic  scheme to the dynamics (\ref{eq:110}) and 
(\ref{eq:321}). In the first case  (\ref{eq:110}), by replacing  $s_i$ with 
a matrix $A_{s_i}$ we find that the cancellation would occur only if 
\bea
A_1A_1A_0 &=&-\tilde A_1 \tilde A_1A_0 + A_1 \tilde A_1\tilde A_0\cr
 &=&\tilde A_1 \tilde A_0A_1 - A_1 \tilde A_0\tilde A_1.
\label{eq:A}  
\eea  

Note, that  these algebraic relations  can not be satisfied by 
non-zero scalars $A_0$ and $A_1$, but  there are solutions  
where $A_0, A_1$ and the auxiliaries are finite 
dimensional matrices\cite{matrix}.

Next, for the dynamics  (\ref{eq:321})  with bond variables
we replace  $\tau_i$ by matrices $X_{\tau_i}$ and demand
that the generic cancellation scheme (\ref{eq:cancel}) should hold 
for this dynamics (\ref{eq:321}). Then, the matrices $\{X_\tau\}$  and auxiliaries  
$\{\tilde X_\tau\}$ must satisfy  
\bea
X_3X_2X_1 &=&  \tilde{X_2}\tilde{X_1}X_3 - X_2\tilde{X_1}\tilde{X_3} \cr
          &=& -\tilde{X_3}\tilde{X_2}X_1 + X_3\tilde{X_2}\tilde{X_1}\label{eq:rd2}\\
X_3X_2X_0 &=&  \tilde{X_2}\tilde{X_1}X_2 - X_2\tilde{X_1}\tilde{X_2} \cr
          &=& -\tilde{X_3}\tilde{X_2}X_0 + X_3\tilde{X_2}\tilde{X_0}\label{eq:rd4}
\eea
It is not difficult to see  that  Eqs. (\ref{eq:rd2}) and  (\ref{eq:rd4}) have  a scalar solution 
$X_0=0, X_1 = X_2= X_3=1 $ with   auxiliaries $\tilde {X_0}=0,\tilde {X_1}=2=\tilde {X_2}$,
$\tilde {X_3}={\frac 3 2}$.  
Usually
the solutions of MPA with one of the matrix being zero  
are not acceptable as it indicates that  certain configurations  are 
never visited in steady state. For example, here $X_0=0$ would mean that steady state 
weight is zero for all  configurations having two or more consecutive zeros.  
Thus, to accept  solutions with $X_0=0$ we must show  {\it a priori} that the steady state
of the above said configurations are \textit{in fact} zero. A direct proof is lengthy.
Alternatively, one can prove the same using 
a mapping of the model to zero range process (ZRP) which is discussed 
later [below Eq.(\ref{eq:f})]. 
Since it is easy to work  with scalars, we choose to continue with dynamics (\ref{eq:321}).
Let us first calculate the partition function keeping in mind 
that (i) $\tau_i$ and $\tau_{i+1}$ have a common  site $s_{i+1}$, 
and (ii) $\sum_i s_i = N$. The first restriction can be taken care of by 
defining  $X_{\tau_i} = \bra s_i |Y| s_{i+1}\ket$. Using the above scalar solution, 
$Y= \left( \begin{array}{cc}0 & 1\cr 1 & 1\end{array} \right).$
Then the partition function is
\beq
Z_{L,N}=  {\sum_{\{s_i\}}}^\prime \prod_i^L \bra s_i |Y| s_{i+1}\ket
\eeq
where $^\prime$ reminds that the  sum is restricted by $\sum_i s_i =N$.   
To evaluate the restricted sum we go over to grand canonical 
system which is  an  ensemble of  $L$ site rings, each having a  weight 
$z^{2N}$ where $N$, the number of particles in the ring, takes all possible values. 
The grand partition function     
\bea
\c Z_L(z)&=& \sum_N^\infty (z^2)^N Z_{LN} 
=   \sum_{\{s_i\}}  \prod_i^L \bra s_i |Y| s_{i+1}\ket z^{s_i+s_{i+1}} \cr
\label{eq:gce1}
&=& Tr(T^L)= \lambda_+^L + \lambda_-^L
\label{eq:gce2}
\eea
where $T= \left( \begin{array}{cc} 0 & z\cr z  & z^2 \end{array} \right),$ and 
 $\lambda_\pm= \frac z 2 (z\pm \sqrt{4+z^2})$ are the eigenvalues of $T$.
Average density of particles  is then
\bea
\rho(z)&=& \lim_{L\to\infty}\frac{\bra N\ket}{L}= z^2\frac{d}{d(z^2)} \ln \c Z \cr 
&=& \lambda_+/(\lambda_+ + \lambda_-),
\label{eq:rho} 
\eea
where $L\to \infty$ limit has been used in the last step. A system with 
fixed density $\rho=N/L$ would correspond to the choice of  $z$ which is 
consistent with Eq.(\ref{eq:rho}), $i.e.$, 
\bea
z={2\rho-1 \over \sqrt{\rho(1-\rho)}}\label{eq:zrho}.
\eea
Now, let us calculate  some of the observables. First, 
the order parameter,  
\bea
\bra \phi_i\ket  &= & {1\over \c Z_L(z)}\bra 1|T|1\ket \bra 1|T|0\ket \bra 0|T^{L-2}|1\ket \cr
&=&\rho_a \left[ \frac{1-\lambda_+^2 (\lambda_- /\lambda_+)^{L-2}}{1+ (\lambda_- /\lambda_+)^{L}}\right]\cr
{\rm where} \n \\
 \rho_a&=&  (2\rho-1)(1-\rho)/\rho\label{eq:rhoam1}
\eea
 is the order parameter of the system  in  the thermodynamic 
limit $L\to \infty$. As $\rho_a=0$ at $\rho=\rho_c=1/2$, 
$\bra \phi_i\ket$  is independent of $L$ at the critical point. However 
for $\rho>\rho_c$,  $ \bra \phi_i\ket $ has a finite size correction and it  
converges to $\rho_a$ exponentially with $L$.

 Any observable can be calculated from the generic  $(n+1)$-point correlation 
function, 
\bea
C_n=\bra s_is_{i+1}...s_{i+n} \ket &=& \lim_{L\to\infty} 
{1\over \c Z_L(z)}\bra 1|T|1\ket^n \bra 1|T^{L-n}|1\ket \n\\
                      &=&\rho \left[{2\rho-1 \over \rho}\right]^n.
\eea
For example, $\bra \phi_i\ket =\bra s_i s_{i+1} (1-s_{i+2}) \ket = C_1 -C_2 =(2\rho-1)(1-\rho)/\rho$, which is same as Eq.(\ref{eq:rhoam1}).
Now, assuming translational invariance, correlation between two active sites separated by 
$j$ lattice sites, $\Gamma_{j}=\bra \phi_i \phi_{i+j}\ket-\bra \phi_i \ket \bra \phi_{i+j}\ket$ can be calculated as follows.
We have, 
\bea
              \bra \phi_i \phi_{i+j}\ket &=&{1 \over \c Z_L}\bra 1|T|1 \ket ^2
\bra 1|T|0 \ket ^2\bra 0|T^{j-2}|1 \ket \bra 0|T^{L-j-2}|1\ket \n \\
  &=&{\rho_a}^2\left[1-\left({1-\rho \over \rho}\right)^{j-2}\right]\nonumber
\eea
resulting in
\bea
\Gamma_j(\rho)=-(2\rho-1)^2\left[{\rho-1\over\rho}\right]^j.
\label{eq:gamma}
\eea
Note that $\Gamma_j(\rho)$ oscillates with $j$ as shown in the inset 
of Fig. 1. Such an oscillation is expected as $\phi_i \phi_{i\pm1}=0=\phi_i \phi_{i\pm2} $ for every $i$.
 
Let us  calculate the critical exponents  of this 
phase transition. 
Formally  the correlation function is written as 
$\Gamma(j)\sim exp(-j/\xi)j^{(-D+2-\eta)}$. Thus, Eq.(\ref{eq:gamma}) implies  that 
$\eta=1$ and that  the correlation length  $\xi= (\ln \frac{\rho}{1-\rho})^{-1}$ 
diverges  as $\xi \sim (\rho-\rho_c)^{-\nu}$ with $\nu=1$. 
From (\ref{eq:rhoam1}), $\rho_a$ is linear in $(\rho-\rho_c)$ near the 
critical point. So, the order parameter exponent $\beta=1$. 
Again, the survival  probability $\c P$ that 
a single active site survives in $t\to\infty$ limit, vanishes as 
$\c P = (\rho -\rho_c)^{\beta'}$. In RASEP, the activity  
certainly survives for any density $\rho > \rho_c$; thus, $\beta'=0.$

\begin{table}[h]
\caption{Critical exponents of DP, CDP, and RASEP}
\begin{tabular}{ccccc}
\hline\hline
 &$\beta$&$\eta$&$\nu$&$\beta^\prime$\\ \hline
DP\cite{dp} & 0.276486 & 1.504144& 1.0968 &0.276486 \\
 
CDP\cite{dp} & 0&1 &1&1\\  
RASEP&1&1&1&0\\ \hline\hline
\end{tabular}
\end{table}
For comparison we have listed all these critical exponents of the model 
along with those for other known  universality classes of active-absorbing 
phase transitions in Table I. The most well-known and generic  
universality class of active-absorbing phase transition having a fluctuating 
scalar order parameter is DP\cite{gras}. Models where order parameters 
obey  special conservation laws could differ from DP. One such example 
is compact directed percolation (CDP)\cite{cdp} where the activity field 
satisfies the particle hole symmetry. In RASEP, the order parameter which is  
scalar and fluctuating, does not satisfy any special conservation law. 
That, it shows an active-absorbing phase transition  different from DP is 
surprising.  Coupling of this fluctuating order parameter to a conserved 
field, namely density,  could be a possible cause. In fact, it is well known
\cite{sp} that in 
sand-pile models of self organized critically,  the activity field (which is 
scalar and fluctuating) is coupled to the conserved height field   
resulting in universality classes different from DP.

To know if this new universality class is  stable against perturbations 
we have studied several variations of the model by introducing stochasticity 
both in  the direction and rate of particle transfer,
\beq
110 \to  101  ~~~ {\rm  and }~ ~ 011 \mathop {\to}^p 101 \\
\label{eq:new}
\eeq
Naturally, here a site  is called active when $s_i=1$ and either of $s_{i\pm1}=0$.
Note that  CLG  in one dimension  is a special case of Eq. (\ref{eq:new}) 
with  $p=1$ where both forward and backward hopping of particles are allowed. 
Since this  symmetric dynamics satisfies detailed balance\cite{Oliveira},  
all the allowed configurations  have the same weight in the steady state . 
For generic $p\ne 1$, however, the dynamics  (\ref{eq:new}) does not satisfy
detailed balance. To obtain the  exact steady state distribution for arbitrary  
$0\le p\le 1$ we use  MPA for three site dynamics described in this paper.  
Here again,    configurations with two or more consecutive zeros are never visited in 
the steady state, and all other configurations are equally  probable.
The transition occurs at the critical density $\rho_c=\frac{1}{2}$.  
The density of active sites in the  active phase ($\rho>\rho_c$) is given by  
$\rho_a= 2(1-\rho)(2\rho-1)/\rho$, which vanishes 
linearly as $\rho\to\rho_c$ resulting in $\beta=1$. Other critical exponents  
$\nu=\eta=1$ are found to be the same as  that of RASEP.  

It is worth mentioning that 
a transformation $1\leftrightarrow0$ of  (\ref{eq:new}) which leads to a dynamics  
\beq
001 \to  010  ~~~ {\rm  and }~ ~ 100 \mathop {\to}^p 010 \\
\label{eq:new2}
\eeq
also shows a transition at $\rho_c=1/2$,  with order parameter
$\rho_a= 2\rho(1-2\rho)/(1-\rho)$ for $\rho<1/2$. The critical behaviour here, as expected,
is same as that  of RASEP. An interesting variation is when  both 
the dynamics  (\ref{eq:new}) and  (\ref{eq:new2})are present. In this case  we have only 
two absorbing states $\{101010\dots\}$ and  $\{010101\dots\}$ which are symmetric.
This may lead to different critical behaviour\cite{sym}, as supported by the 
numerical studies of these models for $p=1$ \cite{park}.

In another variation, a particle from an occupied site hops to its right
only when it is followed by $\mu$ occupied sites from its left.  
Thus, Eq.(\ref{eq:110}) is a special case with $\mu=1$. 
For finite $\mu=2,3 \dots $ the dynamics  are     
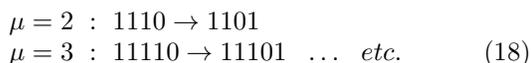
\bea
\mu=2 &:& 1110 \rightarrow 1101\cr
\mu=3 &:& 11110 \rightarrow 11101  ~~\dots~~etc.
\label{eq:mu}
\eea
To use MPA for these dynamics, we have  extended our formulation for 
$(\mu+2)$-site dynamics. The exact results there 
show that this class of models  with $\mu\ge2$  undergo an active 
absorbing phase transition which belong to the  same
universality class  as the system with  $\mu=1$. 
Details of these calculations will be published elsewhere.
Here we  show  a mapping of
these models to the ZRP\cite{zrp} which 
simplify calculations of some of the observables, such as 
$\rho_a$ and its distribution.

The ZRP is defined on a periodic one dimensional lattice with the 
following dynamics; a single particle from a randomly chosen site 
(or box) hops to  one of its neighbour, say  the right one, with  rate 
$u(n)$  which depends on the number of particles $n$ in departure box. 
To map Eq.(\ref{eq:mu}) to ZRP, we define the vacant sites as boxes, 
and the number of uninterrupted  sequence of $1$s  to the left 
of  a vacant site as the number of particles in that box.  
Thus  there are  $N$ particles  which are distributed  among 
$M=L-N$ boxes. 
Now, dynamics (\ref{eq:mu})  just transfers  a 
particle  from a box to its right if the departure box has more 
than $\mu$ particles. Thus,
\beq
u(n) = \theta(n-\mu),
\eeq 
 where $\theta(x)$ is the Heaviside theta function. 
Clearly, when the particles per box $\varrho=\frac NM<\mu$, 
the system has at least one configuration where every box contains $\le\mu$ particles. 
Such configurations are absorbing and   the  system  is arrested 
in one of them in the steady state.  
Thus the critical density is $\varrho_c=\mu$, which 
corresponds to $\rho_c= \frac{\varrho_c}{1+\varrho_c}=\frac {\mu}{1+\mu}$.

Notice  that the steady 
state  weight  in the active phase  $\rho>\rho_c$ has product measure :
\beq 
P(n_1,n_2\dots n_M) \sim  f(n_1) f(n_2) \dots f(n_M),
\label{eq:product}
\eeq
where $n_i$ is the number of particles in box $i$ and function $f$ is to be determined
such that Eq.(\ref{eq:product}) satisfies the master equation in steady state\cite{zrp}. 
In this case, the rate of transfer is independent of number of particles
resulting in
\beq
f(n) = \theta(n+1-\mu).
\label{eq:f}
\eeq
Thus, in steady state, all configurations with {\it every box} containing 
$\mu$ or more particles are visited with equal probability and all other 
configurations, which have  at least one box containing  less than $\mu$  
particles, are  never visited. In particular for RASEP (\ref{eq:110}), 
configurations having two or more consecutive zeros are not allowed 
in steady state (as claimed earlier).

Partition function of the system, in this case, is just the total number of 
configurations where $N$ particles are distributed in $L-N$ boxes such that 
each contains at least $\mu$ particles.  
\bea
Z_{L,N}=  C_{L-N-1}^{N-(\mu-1)(L-N)-1},  
\label{ZLN}
\eea
Note that every configuration $\{s_i\}$ has $L$ translationally equivalent configurations, 
whereas in ZRP it has only only $L-N$ equivalent ones.  
This raises a multiplicative factor $L/(L-N)$ to the steady state weight of {\it every} configuration. We have ignored this factor in Eq.(\ref{ZLN}) and in further calculations 
as it does not affect the observables.

   This mapping allows the calculation of fluctuations in the number  
of active sites $ N_a = \sum_{i=1}^L  \phi_i =\sum_{i=1}^M \theta (n_i -\mu)$.
In a system of $N$ particles distributed among $M$ boxes
probability of finding     
$N_a$ boxes which have more than $\mu$ particles with a restriction that 
every box contains at least $\mu$ particles is given by
a hypergeometric distribution,
\bea
P(N_a)={1\over Z_{L,N}}\; C_{N_a-1}^{N-\mu(L-N)-1}\; C_{N_a}^{L-N}.  
\eea
Mean and variance of this distribution are related to the 
order parameter and  its fluctuation  respectively as
\bea
\rho_a &=& \lim_{L\to \infty} \frac{\bra N_a\ket}{L} =\frac{[\rho-\mu (1-\rho)](1-\rho)}{\rho-(\mu-1)(1-\rho)}\label{eq:rhoamu}\\
\Delta \rho_a &=&  \lim_{L\to \infty}\frac 1L(\bra N_a^2\ket  - \bra N_a\ket^2) \cr 
&=& \frac{{\rho_a}^2}{\rho-(\mu-1)(1-\rho)}
\label{eq:sqrhoa}
\eea
Equation (\ref{eq:rhoamu}) provides  an exact expression of 
$\rho_a$ for the generic model, which  vanishes linearly as 
$\rho$ approaches $\rho_c= \frac{\mu}{1+\mu}$. From  Eq.(\ref{eq:sqrhoa})
it is clear that the fluctuation vanishes quadratically
as $\rho\to \rho_c$. Contrary to  other known  continuous 
transitions, here the transitions are not associated with  
diverging fluctuation of the order parameter  
but it is associated with a diverging correlation length.

Of course,  one  can extend these models to incorporate particle 
transfer to both directions.  A model of ZRP having  a generic  threshold $\mu$ 
and unbiased particle transfer has been studied earlier\cite{Kavita} in  
$d$-dimensions. Characteristic  critical exponets, in the context of 
sand-pile models, have been  discussed.

In conclusion, we have introduced  
a class of models in one dimension where a particle can move to  a vacant 
neighbouring site in one direction only  if 
it is followed by $\mu$ number of particles in the other direction. 
We extended the matrix product ansatz to generic three site dynamics  and 
apply the formalism to  the simplest version of the model with $\mu=1$ to get the
exact steady state distribution. We show that these models 
undergo  a continuous active-absorbing phase transition when density of 
particles is decreased  below $\rho_c= \mu/(1+\mu)$.   
Interestingly  the fluctuation of the order parameter here does not show any 
divergence at the transition point, whereas active sites are found to be correlated 
within a length scale  $\xi$ which diverges  as critical density is approached 
from above.  Critical exponents  of this active-absorbing phase transition,  
which are calculated analytically, are found to be different from the  generic 
universality class, namely directed percolation. We argue that  
the fluctuating scalar order-parameters in these models are  coupled to 
the density field which is conserved, which could be a possible reason why these 
models differ from the  DP universality class. 

We thank D. Dhar for his useful comments.
UB would like to acknowledge thankfully the financial support of the Council of Scientific and
Industrial Research, India (Grant No.SPM-07/489(0034)/2007).

\end{document}